\documentclass[a4paper,12pt]{article}
\pdfoutput=1 % if your are submitting a pdflatex (i.e. if you have
\usepackage{graphicx}
\usepackage{ytableau}
\usepackage{tikz}
\usepackage{jheppub} % for details on the use of the package, please
\newcommand{\be}{\begin{equation}}
\newcommand{\eeq}{\end{equation}}
\newcommand{\bea}{\begin{eqnarray}}
\newcommand{\eea}{\end{eqnarray}}
\newcommand{\ba}{\begin{array}}
\newcommand{\ea}{\end{array}}

\newcommand{\ee}{\end{equation} }

\newcommand{\one}{{\rm 1\kern -.9mm l}}

\newcommand\defeq{\mathrel{\overset{\makebox[0pt]
{\mbox{\normalfont\tiny\sffamily def}}}{=}}}

\title{VEV of Baxter's Q-operator in
N=2 gauge theory and the BPZ differential
equation}

\author{Gabriel Poghosyan}
\author{and Rubik Poghossian}

\affiliation{Yerevan Physics Institute,\\
Alikhanian Br. 2, AM-0036 Yerevan, Armenia}
\emailAdd{gabrielpoghos@gmail.com}
\emailAdd{poghos@yerphi.am}

\abstract{In this short notes using AGT correspondence
we express simplest fully degenerate primary fields of
Toda field theory in terms an analogue of Baxter's $Q$-operator
naturally emerging in ${\cal N}=2$ gauge theory side.
This quantity can be considered as a generating
function of simple trace chiral operators constructed
from  the scalars of the ${\cal N}=2$ vector multiplets.
In the special case of Liouville theory, exploring
the second order differential equation satisfied by conformal
blocks including a degenerate at the second level primary field
(BPZ equation) we derive a mixed difference-differential
relation for $Q$-operator. Thus we generalize the $T$-$Q$ difference equation known in Nekrasov-Shatashvili limit
of the $\Omega$-background to the generic case.
}


\preprint{YerPhI/2016/02}
\begin{document}

\maketitle
\flushbottom
\section{Introduction}
Instanton \cite{Belavin:1975fg} partition function of ${\cal N}=2$ supersymmetric gauge
theory in $\Omega$-background admits exact investigation by
localization methods
\cite{Lossev:1997bz,Nekrasov:2002qd,Flume:2002az,
Bruzzo:2002xf,Nekrasov:2003rj}.
In the limit
when the background parameters $\epsilon_1$, $\epsilon_2$ vanish,
the famous Seiberg-Witten solution
\cite{Seiberg:1994aj,Seiberg:1994rs}
is recovered. The case of non-trivial $\Omega$-background has surprisingly rich area of applications. In
particular when one of parameters is set to zero
(Nekrasov-Shatashvili limit \cite{Nekrasov:2009rc}), deep relations to
quantum integrable system emerge (see e.g. \cite{Mironov:2009uv,Mironov:2009dv,Maruyoshi:2010iu,
Poghossian:2010pn,Fucito:2011pn,Fucito:2012xc,
Nekrasov:2012xe,Nekrasov:2013xda}
to quote a few from many important works). These are
quantum versions of classical
integrable systems, which played central role already in Seiberg-Witten theory on trivial background \cite{Gorsky:1995zq,Martinec:1995by}. The remaining non-zero
$\Omega$-background parameter just plays the role of Plank's constant.
Many familiar concepts of exactly integrable models of
statistical mechanics and quantum field theory such as
Bethe-ansatz or Baxters $T-Q$ equations
\cite{Baxter:1982,Bazhanov:1998dq} naturally emerge
in this context \cite{Poghossian:2010pn}. In the case of generic $\Omega$-background
instanton partition function is directly related to the conformal
blocks of a 2d CFT (AGT correspondence)
\cite{Alday:2009aq,Wyllard:2009hg,Poghossian:2009mk,
Alba:2010qc,Fateev:2011hq}.
In this context the NS limit corresponds to the semi-classical limit
of the related CFT
\cite{Alday:2009fs,Maruyoshi:2010iu,Piatek:2011tp,
Nekrasov:2013xda,Piatek:2013ifa,Ashok:2015gfa,Poghossian:2016rzb}.

In \cite{Poghossian:2016rzb} one of present authors (R.P.) has investigated the link
between Deformed Seiberg-Witten curve equation and underlying
Baxter's $T-Q$ equation in gauge theory side and the null-vector
decoupling equation \cite{Belavin:1984vu} of 2d CFT in quite general setting
of linear quiver gauge theories with $U(n)$ gauge groups and
2d $A_{n-1}$ Toda field theory multi-point conformal blocks
in semi-classical limit (see also \cite{Marshakov:2010fx} for
an earlier discussion on the role of degenerate fields in
AGT correspondence).

In this short notes we'll extend some of the  results of
\cite{Poghossian:2016rzb}
to the case of generic $\Omega$-background corresponding
to the genuine quantum conformal blocks. For technical
reasons we'll restrict ourselves to the case of $U(2)$ gauge
groups corresponding to the Liouville theory leaving Toda
field theory case for future work.

In Section \ref{Section2} we show that an appropriate
choice of parameters \cite{Fucito:2013fba} in
$A_{r+1}$ linear quiver theory
with $U(n)$ gauge groups is equivalent to insertion of
the analoge of Baxters $Q$-operator into the partition function
of a theory with one gauge node less $A_r$ theory with generic
parameters. In the 2d CFT  side such special choice corresponds
to insertion of a degenerated primary field in the conformal
block \cite{Fucito:2013fba}. In Section \ref{Section2}
restricting to the case of Liouville theory, starting from the
second order differential equation satisfied by the
multi-points conformal blocks including a degenerate field
$V_{-b/2}$ \cite{Belavin:1984vu} we derive the analogues equation satisfied by
the gauge theory partition function with $Q$ operator insertion.
Then we show that this equation leads to a mixed linear
difference-differential equation for $Q$ operators which is
a direct generalization of the $T-Q$ equation from NS limit
to the case of generic $\Omega$-Background. Finally we
summarize our results and discuss a couple of further directions
which we think are worth pursuing.
%%%%%%%%%%%%%%%%%%%%%%%%%%%%%%%%%%%%%%%%%%%%%%%%%%%%%%%%%%%%%%%%%%%%%
%%%%%%%%%%%%%%%%%%%%%%%%%%%%%%%%%%%%%%%%%%%%%%%%%%%%%%%%%%%%%%%%%%%%%
\section{A special choice of parameters, leading to
$\mathbf{Q}_{\vec{Y}}$ insertion}
\label{Section2}
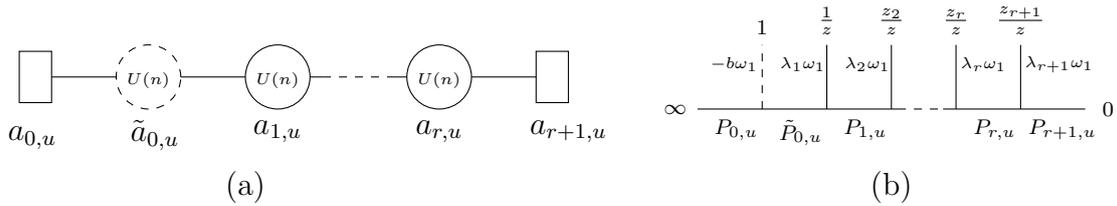
\begin{figure}[t]
\begin{tikzpicture}[scale=0.85]
%\draw[help lines] (0,0) grid (23,1);
% Quiver gauge theory
\draw [] (0.5,0.1) rectangle (1,0.9);
\node [below] at (0.7,-0.1) {$a_{0,u}$};
\draw [] (1,0.5)--(2,0.5);
\draw [dashed] (2.5,0.5) circle [radius=0.5];
\node [] at (2.5,0.45) {\tiny{$U(n)$}};
\node [below] at (2.6,0) {$\tilde{a}_{0,u}$};
\draw [] (3,0.5)--(4,0.5);
\draw [] (4.5,0.5) circle [radius=0.5];
\node [] at (4.5,0.45) {\tiny{$U(n)$}};
\node [below] at (4.5,0) {$a_{1,u}$};
\draw [] (5,0.5)--(5.2,0.5);
\draw [dashed] (5.2,0.5)--(6.3,0.5);
\draw [] (6.3,0.5)--(6.5,0.5);
\draw [] (7,0.5) circle [radius=0.5];
\node [] at (7,0.45) {\tiny{$U(n)$}};
\node [below] at (7,0) {$a_{r,u}$};
\draw [] (7.5,0.5)--(8.5,0.5);
\draw [] (8.5,0.1) rectangle (9,0.9);
\node [below] at (9,0) {$a_{r+1,u}$};
%CFT block
\draw [] (11,0)--(14.1,0);
\draw [dashed] (14.1,0)--(14.8,0);
\draw [] (14.8,0)--(17,0);
\node [left] at (11,0) {\scriptsize{$\infty$}};
\node [below] at (11.6,0) {\scriptsize{$P_{0,u}$}};
\node [below] at (12.6,0) {\scriptsize{$\tilde{P}_{0,u}$}};
\node [below] at (13.6,0) {\scriptsize{$P_{1,u}$}};
\node [below] at (15.6,0) {\scriptsize{$P_{r,u}$}};
\node [below] at (16.65,0) {\scriptsize{$P_{r+1,u}$}};
\node [right] at (17.1,0) {\scriptsize{$0$}};
\draw [dashed] (12,0)--(12,1);
\node [left] at (12.1,0.7) {\tiny{$-b\omega_1$}};
\node [above] at (12,1) {\scriptsize{$1$}};
\draw [] (13,0)--(13,1);
\node [left] at (13.15,0.7) {\tiny{$\lambda_1\omega_1$}};
\node [above] at (13,1) {\scriptsize{$\frac{1}{z}$}};
\draw [] (14,0)--(14,1);
\node [left] at (14.15,0.7) {\tiny{$\lambda_2\omega_1$}};
\node [above] at (14,1) {\scriptsize{$\frac{z_2}{z}$}};
\draw [] (15,0)--(15,1);
\node [left] at (15.95,0.7) {\tiny{$\lambda_{r}\omega_1$}};
\node [above] at (15,1) {\scriptsize{$\frac{z_{r}}{z}$}};
\draw [] (16,0)--(16,1);
\node [right] at (15.9,0.7) {\tiny{$\lambda_{r+1}\omega_1$}};
\node [above] at (16,1) {\scriptsize{$\frac{z_{r+1}}{z}$}};
\node [below] at (14,-0.8) {(\cal{b})};
\node [below] at (4,-0.8) {(\cal{a})};
\end{tikzpicture}
\caption{(\cal{a}) The quiver diagram for the conformal
linear quiver $U(n)$ gauge theory:
$r$ circles stand for gauge multiplets; two squares represent $n$ anti-fundamental (on the left edge) and $n$ fundamental (the right edge) hypermultiplets; the lines connecting adjacent circles are the bi-fundamentals.
(\cal{b}) The AGT dual conformal block of the Toda field theory.}
\label{quiv_block}
\end{figure}

\begin{figure}
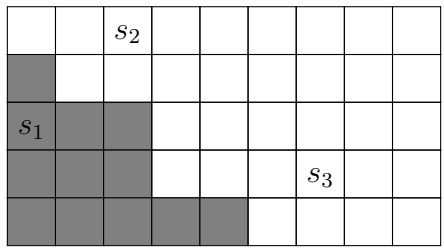

\vspace{0.5cm}
  \centering
\ytableausetup{nosmalltableaux}
\centering
\begin{ytableau}
*(white) &  &s_2 & & & & & &\\
*(gray) &  & & & & & & &\\
*(gray)s_1 & *(gray) &*(gray) & & & & & &\\
*(gray) & *(gray) &*(gray) & & & & s_3& &\\
*(gray) & *(gray) & *(gray)&*(gray) &*(gray)& & & &
\end{ytableau}
\caption{Arm and leg length with respect to a Young diagram
$\lambda=\{4,3,3,1,1\}$ (the gray area): $A_\lambda(s_1)=1$, $L_\lambda(s_1)=2$, $A_\lambda(s_2)=-2$,
$L_\lambda(s_2)=-3$, $A_\lambda(s_3)=-2$, $L_\lambda(s_3)=-4$.}
\label{Arm_Leg}
\end{figure}

Consider the instanton partition function of the linear
quiver theory $A_{r+1}$ with gauge groups $U(n)$ with parameters
specified as in Fig.\ref{quiv_block}a. Note that
the parameters of the first
gauge factor (depicted as a dashed circle) are chosen to be
$\tilde{a}_{0,u}=a_{0,u}-\epsilon_1\delta_{1,u}$,
where $a_{0,u}$ are the parameters
of the "frozen" node corresponding to the $n$ antifundamental
hypermultiplets. It has been shown in \cite{Fucito:2013fba} that under such choice
of parameters all $n$-tuples of Young diagrams $Y_{\tilde{0},u}$ corresponding to the special node $\tilde{0}$ (the dashed circle)
give no contribution in partition function unless the first
diagram $Y_{\tilde{0},1}$ consists of a single column while the remaining
$n-1$ diagrams are empty. Taking into account this huge simplification
we'll be able to separate the contribution of the special node
explicitly. According to the rules of construction of the partition
function for this contribution we have
\bea
\prod_{u,v=1}^n\frac{Z_{bf}\left(a_{0,u},\varnothing|
\tilde{a}_{0,v},Y_{\tilde{0},v}\right)
Z_{bf}\left(\tilde{a}_{0,u},Y_{\tilde{0},u}
|a_{1,v},Y_{1,v}\right)}
{Z_{bf}\left(\tilde{a}_{0,u},Y_{\tilde{0},u}
|\tilde{a}_{0,v},Y_{\tilde{0},v}\right)}
\label{dashed_circle_cont}
\eea
where for a pair of Young diagrams $\lambda$, $\mu$ the bifundamental
contribution is given by
\bea
\label{Zbf}
&&Z_{bf}(a,\lambda|b,\mu)=\\
&&\qquad\quad\prod_{s\in \lambda}
(a-b-\epsilon_1 L_\mu(s)+\epsilon_2 (1+A_\lambda(s)))
\prod_{s\in \mu}
(a-b+\epsilon_1 (1+L_\lambda(s))-\epsilon_2 A_\mu(s)),\nonumber
\eea
where the arm and leg lengths of a box $s$ $A_\lambda(s)$
and $L_\lambda(s)$ towards a Young diagram $\lambda$
are defined as
\bea
A_\lambda(s)=\lambda_i-j\,;\qquad L_\lambda(s)=\lambda_i^\prime-j\,,
\eea
where $(i,j)$ are coordinates of the box $s$ with respect to the
center of the corner box and $\lambda_i$ ($\lambda_j^\prime$) is
the $i$-th column length ($j$-th row length) of $\lambda$ as shown
in Fig.\ref{Arm_Leg}.
%%%%%%%%%%%%%%%%%%%%%%%%%%%%%%%%%%%%%%%%%%%%%%%%%%%%%%%%%%%%%%%%%%%%%
%%%%%%%%%%%%%%%%%%%%%%%%%%%%%%%%%%%%%%%%%%%%%%%%%%%%%%%%%%%%%%%%%%%%%
%%%%%%%%%%%%%%%%%%%%%%%%%%%%%%%%%%%%%%%%%%%%%%%%%%%%%%%%%%%%%%%%%%%%%%%%%%%

%%%%%%%%%%%%%%%%%%%%%%%%%%%%%%%%%%%%%%%%%%%%%%%%%%%%%%%%%%%%%%%%%%%%%%%%%%%
%%%%%%%%%%%%%%%%%%%%%%%%%%%%%%%%%%%%%%%%%%%%%%%%%%%%%%%%%%%%%%%%%%%%%%%%%%%
Using (\ref{Zbf}) It is not difficult to compute the factors $Z_{bf}$ present
in (\ref{dashed_circle_cont}). In particular
\bea
Z_{bf}\left(a,\varnothing|b,\lambda\right)=
\prod_{s\in\lambda}(a-b-\varphi(s))\,
\label{Zbf_empty}
\eea
where
\bea
\varphi(s)=\epsilon_1(i_s-1)+\epsilon_2(j_s-1)\,
\eea
(e.g. in Fig.\ref{Arm_Leg} $\varphi(s_3)=6\epsilon_1+\epsilon_2$).
To present the final result for the contribution (\ref{dashed_circle_cont}) it is convenient to introduce the notation
\bea
\mathbf{Q}(v|\lambda)=\frac{(-\epsilon_2)^{\frac{v}{\epsilon_2}}}
{\Gamma(-\frac{v}{\epsilon_2})}\prod_{s\in\lambda}
\frac{v-\varphi(s)+\epsilon_1}{v-\varphi(s)}
\label{single_Q}
\eea
The analogues quantity  was
instrumental in construction of Baxters T-Q relation in the context of
Nekrasov-Shatashvili limit of ${\cal N}=2$ gauge theories \cite{Poghossian:2010pn}. Recently the
importance of this quantity in the case generic $\Omega$-background was
emphasized in \cite{Nekrasov:2015wsu}.
A careful examination shows that the contribution
(\ref{dashed_circle_cont}) can be conveniently represented as
\bea
\prod_{u=1}^n\,\frac{\mathbf{Q}\left(a_{0,1}-a_{1,u}+\epsilon_2 k|Y_{1,u}\right)}
{\epsilon_2^k\left(\frac{a_{0,1}-a_{0,u}+\epsilon_2}{\epsilon_2}\right)_k
\mathbf{Q}\left(a_{0,1}-a_{1,u}|Y_{1,u}\right)}\,\prod_{u,v=1}^n
Z_{bf}\left(\tilde{a}_{0,u},\varnothing
|a_{1,v},Y_{1,v}\right)\,
\eea
where
\bea
(x)_k=x(x+1)\cdots (x+k-1)=\frac{\Gamma(x+k)}{\Gamma(x)}
\eea
is the Pochammer's symbol.
Using (\ref{Zbf_empty}) we can see that the Young diagram
dependent part of factor $Q$ in the denominator can be absorbed
in the double product. The net effect
is a simple replacement of parameters $\tilde{a}_{u,0}$
by $a_{u,0}$ in arguments of the functions $Z_{bf}$:
\bea
\prod_{u=1}^n\frac{\Gamma \left(-\frac{a_{0,1}-a_{1,u}}{\epsilon _2}\right)\mathbf{Q}\left(a_{0,1}-a_{1,u}+\epsilon_2 k|Y_{1,u}\right)}
{\epsilon_2^k\left(-\epsilon _2\right){}^{\frac{a_{0,1}-a_{1,u}}{\epsilon _2}}\left(\frac{a_{0,1}-a_{0,u}+\epsilon_2}{\epsilon_2}\right)_k}
\prod_{u,v=1}^n
Z_{bf}\left(a_{0,u},\varnothing
|a_{1,v},Y_{1,v}\right)
\label{dashed_circle_as_insert}
\eea
Thus we conclude that $k$-instanton sector of the dashed circle
in $A_{r+1}$ linear quiver theory can be treated as insertion
of the operator
\bea
\mathbf{Q}_{\vec{Y_1}}(a_{0,1}+k\epsilon_2)=
\prod_{u=1}^n\mathbf{Q}\left(a_{0,1}-a_{1,u}
+\epsilon_2 k|Y_{1,u}\right)
\label{Q_insertion}
\eea
in a generic $A_r$ theory. It was already known \cite{Fucito:2013fba},
that the special
choice of parameters $\tilde{a}_{0,u}=a_{0,u}
-\epsilon_1\delta_{u,1}$ corresponds
to the insertion of the completely degenerate field
$V_{-b\omega_1(z)}$ in AGT dual Toda CFT conformal block.
Thus (\ref{Q_insertion}) gives an explicit realization of this
field in terms of ${\cal N}=2$ gauge theory notions.

Until now we were discussing arbitrary gauge $U(n)$ gauge factors.
In what follows, we'll restrict ourselves with the case $n=2$,
corresponding to the Liouville theory in AGT dual side. The reason
is that in Liouville theory conformal blocks including this
degenerate field, satisfy second order differential equation
\footnote{In generic Toda theory, the analogues null vector
decoupling equation is not investigated in full details yet.
Instead there is a recent progress in the case of
quasi-classical limit \cite{Poghossian:2016rzb}.}.
In remaining part of the
paper we'll translate this differential equation in gauge
theory terms, finding a linear difference-differential equation, satisfied by the expectation values of the operators
$\mathbf{Q}(v)$. Since the equation is valid for infinitely
many discrete values of the spectral parameter
$v=a_{0,1}+k \epsilon_2$, $k=0,1,2,\ldots $, it can be
argued that it is valid for generic values of $v$ as well.
The last statement we have checked
also by explicit low order instanton computations.
%%%%%%%%%%%%%%%%%%%%%%%%%%%%%%%%%%%%%%%%%%%%%%%%%%%%%%%%%%%%%%%%%%%%%%%%%%%
\section{Degenerate field decoupling equation in Liouville theory}
\label{Section3}
Let us briefly remind that the Liouvill theory (see e.g. \cite{Zamolodchikov:1995aa}) is characterized
by the central charge $c$ of Virasoro algebra parameterized as
\bea
c=1+6 Q^2\,\qquad Q=b+\frac{1}{b}\,
\eea
where $b$ is the Liouvill's dimensionless coupling constant
related to the $\Omega$-background parameters
via
\bea
b=\sqrt{\epsilon_1\over\epsilon_2}
\eea
The conformal dimensions of primary fields
are $V_\lambda$ are given by
\bea
h(\lambda)=\lambda(Q-\lambda)\,.
\eea
The parameters $\alpha$ are usually referred as charges. One
alternatively uses the Liouville momenta $P=Q/2-\lambda$.
In Fig.\ref{quiv_block}b we found it convenient to specify
the fields associated to the horizontal lines by their
momenta, while those of vertical lines by charges. The relations
between this parameters and the gauge theory VEV's are very
simple\footnote{The reader should be careful, there are
various factors of $2$ between specialized to $n=2$ Toda notations
compared to the standard Liouville theory conventions, adopted
also in this paper.}
\bea
p_{\alpha}=\frac{1}{\sqrt{\epsilon_1\epsilon_2}}
\frac{a_{\alpha,1}-a_{\alpha,2}}{2}\,;\qquad \lambda_\alpha=\frac{1}{\sqrt{\epsilon_1\epsilon_2}}
\left(\frac{a_{\alpha,1}+a_{\alpha,2}}{2}-
\frac{a_{\alpha-1,1}+a_{\alpha-1,2}}{2}\right)
\label{AGT_map_1}
\eea
for $\alpha=2,3,\ldots,r+1$. With the same logic we have
\bea
&&p_{0}=\frac{1}{\sqrt{\epsilon_1\epsilon_2}}
\frac{a_{0,1}-a_{0,2}}{2}\,;\,\,\quad
p_{\tilde{0}}=\frac{1}{\sqrt{\epsilon_1\epsilon_2}}
\frac{a_{0,1}-\epsilon_1-a_{0,2}}{2}\nonumber\\
&&\lambda_{\tilde{0}}=-\frac{\epsilon_1}{\sqrt{\epsilon_1\epsilon_2}}
=-\frac{b}{2}\,;\qquad\lambda_1=
\frac{\epsilon_1}{\sqrt{\epsilon_1\epsilon_2}}
\left(\frac{a_{1,1}+a_{1,2}}{2}
-\frac{a_{0,1}-\epsilon_1+a_{0,2}}{2}\right)
\label{AGT_map_2}
\eea
Notice that the field $V_{\lambda_{\tilde{0}}}=V_{-b/2}$ is indeed
a degenerate field satisfying second order differential
equation due to the null vector decoupling condition
(below $L_m$ are the Virasoro generators)
\bea
(b^{-2}L_{-1}^2+L_{-2})V_{-b/2}=0
\eea
The differential equation satisfied by our $r+4$-point
conformal block
\bea
G(z|z_\alpha)=\langle p_0|V_{-b/2}(z)
V_{\lambda_1}(1)V_{\lambda_2}(z_2)\cdots
V_{\lambda_{r+1}}(z_{r+1})|p_{r+1}
\rangle_{\{\tilde{p}_0,\ldots,p_r\}}
\label{CFT_block}
\eea
reads \cite{Belavin:1984vu}
\bea
\left(b^{-2}\partial_z^2-\frac{2 z-1}{z (z-1)}\,\partial_z
+\frac{\delta}{z(z-1)}
+\sum_{\alpha=2}^{r+1}\frac{z_\alpha \left(z_\alpha-1\right)}{z (z-1) \left(z-z_\alpha\right)}\,\partial_{z_\alpha}\,\,\quad\right.\nonumber\\
\left.+\sum_{\alpha=1}^{r+2} \frac{h(\lambda_\alpha)}{(z-z_{\alpha})^2}
\right)G(z|z_\alpha)=0\,
\label{diff_eq_G}
\eea
where
\bea
\delta=h\left(Q/2-p_0\right)-h\left(-b/2\right)
-\sum_{\alpha=1}^{r+2}h(\lambda_\alpha)\quad \text{and}\quad \lambda_{r+2}=Q/2-p_{r+1}\,.
\eea
According to AGT correspondence the instanton part of the
partition function of the ${\cal N}=2$ theory considered
in previous section with $U(2)$ gauge group factors is related
to the conformal block (\ref{CFT_block}) as
\bea
&&G(z|z_\alpha)=Z_{inst}\,\,z^{h(Q/2-p_0)-h(-b/2)
-b\sum_{\alpha=1}^{r+1}(Q-\lambda_\alpha)}
\prod_{\alpha=1}^{r+1}\left(z-z_\alpha\right)
^{b(Q-\lambda_\alpha)}\nonumber\\
&&\times\prod_{1\le \alpha<\beta\le r+1}\left(z_\alpha-z_\beta\right)
^{-2\lambda_\alpha(Q-\lambda_\beta)}
\prod_{\alpha=2}^{r+1}z_\alpha
^{p_\alpha^2-p_{\alpha-1}^2-h(\lambda_\alpha)
+2\lambda_\alpha\sum_{\beta=\alpha+1}^{r+1}(Q-\lambda_\beta)}.
\label{Z_vs_block}
\eea
To complete the map (\ref{AGT_map_1}), (\ref{AGT_map_1})
between two sides let us mention also
that the exponentiated gauge couplings (instanton
counting parameters) are related to the insertion
points as \cite{Alday:2009aq}
\bea
q_\alpha=z_{\alpha+1}/z_\alpha\,;\quad \text{for}
\quad \alpha=1,\ldots,r\,,
\eea
the remaining coupling associated to the special node
$\tilde{0}$ is just $1/z$ and $z_1=1$.

In (\ref{Z_vs_block}) besides standard AGT $U(1)$ factors
an extra power of $z$ responsible for scale transformation
(with scaling factor $z$) mapping the insertion points shown in Fig.\ref{quiv_block}b to those of the conformal block
(\ref{CFT_block}). Inserting (\ref{Z_vs_block}) into
(\ref{diff_eq_G}) and replacing CFT parameters by their
gauge theory counterparts we'll find a differential equation
satisfied by the partition function. After tedious
but straightforward transformations it is possible to represent
this equation as (for more details on calculations of this
kind see \cite{Poghossian:2016rzb})
\bea
\sum_{\alpha=0}^{r+1}(-)^\alpha
\chi_\alpha(-\epsilon_2\,z\partial_z;\hat{u}_1,\ldots,\hat{u}_{r+1})
z^{-\alpha-a_{0,1}/\epsilon_2}Z_{inst}=0
\label{diff_eq}
\eea
where
\bea
\hat{u}_1=-\epsilon_1\epsilon_2\sum_{\alpha=2}^{r+1}
z_\alpha\partial_{z_\alpha}\,;\qquad
\hat{u}_\alpha=\epsilon_1\epsilon_2
z_\alpha\partial_{z_\alpha} \qquad \text{for}
\qquad \alpha=2,\ldots,r+1
\eea
and
$\chi_\alpha(v;u_1,\ldots,u_{r+1})$ are quadratic in
$v$ and linear in $u_1,\ldots,u_{r+1}$ polynomials
(we use notation $\epsilon=\epsilon_1+\epsilon_2$)
\bea
\chi_{\alpha}(v;u_1,\ldots,u_{r+1})=
\sum_{1\le k_1<\cdots<k_{\alpha}\le r+1}
\bigg(\prod_{\beta=1}^\alpha z_{k_\beta}\bigg)
\bigg( y_0(v+\alpha \epsilon
+(\alpha-\delta _{k_1,1})\epsilon_1)\nonumber\\
-\sum _{\beta=1}^\alpha
\left(y_{k_\beta-1}(v+(\alpha-\beta+1)\epsilon
+(\alpha-\delta_{k_1,1})\epsilon_1)
-y_{k_\beta}(v+(\alpha-\beta)\epsilon
+(\alpha-\delta_{k_1,1})\epsilon_1)\right.\nonumber\\
\left. +u_{k_\beta}
+(c_{0,1}-c_{k_\beta-1,1})(c_{k_\beta-1,1}
-c_{k_\beta,1})\right)\hspace{1.3cm}\nonumber\\
\left.+\sum_{1\le\beta<\gamma\le \alpha}(c_{k_\beta-1,1}
-c_{k_\beta,1})(c_{k_\gamma-1,1}
-c_{k_\gamma,1})\right),\hspace{1cm}
\eea
where for $\alpha=0,1,\ldots,r+1$
\bea
y_\alpha(v)=(v-a_{\alpha,1})(v-a_{\alpha,2})\defeq
v^2-c_{\alpha,1}v+c_{\alpha,2}\,.
\eea
We set by definition
\bea
\chi_0(v)=y_0(v)
\eea
and for the other extreme value $\alpha=r+1$ it is
easy to see that
\bea
\chi_{r+1}(v)=y_{r+1}(v)\,\prod_{\beta=1}^rz_\beta\,.
\eea
Representing $Z_{inst}$ as a power series in $1/z$ ,
\bea
Z_{inst}=\sum_{v\in a_{0,1}+\epsilon_2 \mathbb{Z}}
Q(v)z^{-(v-a_{0,1})/\epsilon_2}
\eea
from eq. (\ref{diff_eq}) for the coefficients $Q(v)$ we get
the relation
\bea
\sum_{\alpha=0}^{r+1}(-)^\alpha
\chi_\alpha(v;\hat{u}_1,\ldots,\hat{u}_{r+1})
Q(v-\alpha\epsilon_2)=0\,,
\label{diff_difference}
\eea
 which is valid for infinitely many values $v\in a_{0,1}+\epsilon_2 \mathbb{Z}$.
Since $Z_{inst}$ is regular at $z=\infty$, in fact we have
nontrivial equations only for $v_k=a_{0,1}+\epsilon_2 k$, with
$k\ge 0$.

Remind now that as discussed in previous section, due to eqs. (\ref{dashed_circle_as_insert}), (\ref{Q_insertion}), $Z_{inst}$
of the $A_{r+1}$ theory up to a simple factor is the  same as
VEV of the quantity $\mathbf{Q}_{\vec{Y_1}}$
(\ref{Q_insertion}) calculated in the framework of $A_r$
gauge theory (i.e. in theory without the dashed circle in
Fig.\ref{quiv_block}a). Explicitly
\bea
Q(v_k)=C
\prod_{u=1}^2\frac{\epsilon_2^{(a_{0,1}-v_k)
/\epsilon_2}}{\Gamma\left(\frac{v_k-a_{0,u}}
{\epsilon_2}+1\right)}\,\,\langle
\mathbf{Q}_{\vec{Y}_{1}}\left(v_k\right)
\rangle_{A_r}\,,
\label{Q_VEV}
\eea
where the constant $C$ takes the value
\bea
C=\prod_{u=1}^2\frac{\Gamma
\left(\frac{a_{1,u}-a_{0,1}}{\epsilon _2}\right)
\Gamma\left(\frac{a_{0,1}-a_{0,u}}{\epsilon _2}+1\right)}
{\left(-\epsilon _2\right){}^{\frac{a_{0,1}-a_{0,u}}{\epsilon_2}
}}\,,
\eea
if one adopts conventional unit normalization for both
partition function and the conformal block.
The right hand side of this equation can be calculated
by means of gauge theory for arbitrary $v\in \mathbb{C}$.
There are all reasons to believe that also for generic
values of $v$ the equation (\ref{diff_difference}) still holds.
Indeed, for a given instanton order, the equation (\ref{diff_difference}) states, that some combination
of rational functions\footnote{Evidently, by multiplying with
suitable gamma and exponential functions it is easy
to get rid of non-rational prefactors of (\ref{single_Q}),
(\ref{Q_VEV}).}
of $v$ vanish for all values $v=v_k$, but this is possible
only if this combination vanishes identically.

A simple inspection ensures that the equation (\ref{diff_difference})
in Nekrasov-Shatashvili limit completely agrees with the
analogous difference equation investigated in details in \cite{Poghossian:2016rzb}.

\section{Summary}
Thus we made an explicit link between the insertion of
the $\mathbf{Q}$ operator in ${\cal N}=2$ gauge theory
and insertion of simplest degenerate field in AGT dual
2d CFT.

In the special case of the gauge groups $U(2)$
we found analog of the Baxter's $T-Q$ equation, previously
known only in the Nekrasov-Shatashvili limit of the
$\Omega$-background \cite{Poghossian:2010pn,Fucito:2011pn,Fucito:2012xc,
Nekrasov:2012xe,Nekrasov:2013xda}.

To conclude let us mention that a "microscopic" proof of
this statement e.g. along the line presented in
\cite{Bourgine:2015szm} to prove qq-character identities of \cite{Nekrasov:2015wsu} would be highly desirable.

Another important contribution would be generalization
of our analysis to the case of arbitrary $U(n)$ or other
choices of gauge groups.

%%%%%%%%%%%%%%%%%%%%%%%%%%%%%%%%%%%%%%%%%%%%%%%%%%%%%%%%%%%%%%%%%%%
\vspace{1cm}

\centerline{\large\bf Acknowledgments}

\vspace{0.5cm}
This work was supported by the Armenian State Committee of Science
in the framework of the research projects 15T-1C233 and 15T-1C308.
In part of R.P. this work was partially supported also by the
Volkswagen Foundation of Germany. R.P. is grateful to A. Morozov
for a communication on the reference \cite{Marshakov:2010fx}.
%%%%%%%%%%%%%%%%%%%%%%%%%%%%%%%%%%%%%%%%%%%%%%%%%%%%%%%%%%%%%%%%%%%%%%%%     %%%%%%%%%%%%%%%%%%%%%%%%%%%%%%%%%%%%%%%%%%%%%%%%%%%%%%%%%%%%%%%%%%%%%
\newpage
\providecommand{\href}[2]{#2}\begingroup\raggedright\endgroup
\end{document}